\documentclass[12pt]{article}
\pdfoutput=1

\usepackage{amsmath}
\usepackage{amssymb}
\usepackage{epsfig}
\usepackage{epstopdf}
\usepackage{latexsym}
\usepackage{color}
\numberwithin{equation}{section}
\usepackage[
      colorlinks=false,
      linkcolor=darkblue,  
      urlcolor=blue,    
      filecolor=blue,     
      citecolor=red,
linktocpage=true,
      pdfstartview=FitV,
      bookmarksopen=true    
      ]{hyperref}

\begin{document}

\begin{titlepage}

\centerline
\centerline
\centerline
\bigskip
\centerline{\Huge \rm M2-branes wrapped on a topological disk} 
\bigskip
\bigskip
\bigskip
\bigskip
\bigskip
\bigskip
\bigskip
\bigskip
\centerline{\rm Minwoo Suh}
\bigskip
\centerline{\it Department of Physics, Kyungpook National University, Daegu 41566, Korea}
\bigskip
\centerline{\tt minwoosuh1@gmail.com} 
\bigskip
\bigskip
\bigskip
\bigskip
\bigskip
\bigskip
\bigskip

\begin{abstract}
\noindent Employing the method applied to M5-branes recently by Bah, Bonetti, Minasian and Nardoni, we study M2-branes on a disk with non-trivial holonomies at the boundary. In four-dimensional $U(1)^4$-gauged $\mathcal{N}=2$ supergravity, we find supersymmetric $AdS_2$ solutions from M2-branes wrapped on a topological disk in Calabi-Yau two-, three- and four-folds. We uplift the solutions to eleven-dimensional supergravity. For the solutions from topological disk in Calabi-Yau four-folds, the Bekenstein-Hawking entropy is finite and well-defined. On the other hand, from the topological disk in Calabi-Yau two- and three-folds, we could not find solutions with finite Bekenstein-Hawking entropy.
\end{abstract}

\vskip 6cm

\flushleft {September, 2021}

\end{titlepage}

\tableofcontents

\section{Introduction}

A simple but instructive setup for the AdS/CFT correspondence, \cite{Maldacena:1997re}, is provided by the topological twist in field theory, \cite{Witten:1988ze, Bershadsky:1995vm, Bershadsky:1995qy}, and also in supergravity, \cite{Maldacena:2000mw}. On the other hand, currently, we observe new classes of $AdS$ solutions of supergravity theories beyond the topological twist. First, there are $AdS$ solutions from branes wrapped on a spindle which is topologically a two-sphere with orbifold singularities at the poles. The $AdS$ solutions from D3-branes, \cite{Ferrero:2020laf, Hosseini:2021fge, Boido:2021szx}, M2-branes, \cite{Ferrero:2020twa, Cassani:2021dwa}, and M5-branes, \cite{Ferrero:2021wvk}, wrapped on a spindle are recently found. Although the solutions from D3-branes were previously found in \cite{Cvetic:1999xp, Gauntlett:2006af, Kunduri:2006uh, Gauntlett:2006ns, Kunduri:2007qy} in diverse contexts, the interpretation as a spindle and their AdS/CFT correspondence are new. See also \cite{Couzens:2021tnv}. 

Secondly, there are $AdS$ solutions from branes wrapped on a topological disk which is a disk with non-trivial $U(1)$ holonomies at the boundary. The $AdS_5$ solutions from M5-branes wrapped on a topological disk was found in \cite{Bah:2021mzw, Bah:2021hei}. The dual field theory was proposed to be the Argyres-Douglas theory, \cite{Argyres:1995jj}, from 6d $\mathcal{N}=(2,0)$ theories on a sphere with irregular punctures. The construction was soon applied to $AdS_3$ solutions from D3-branes and M5-branes in \cite{Suh:2021ifj} and $AdS_4$ solutions from D4-D8-brane system, \cite{Suh:2021aik}. To recapitulate, topological disk and spindle are not manifolds with constant curvature and the supersymmetry is not realized by topological twist. 

In this paper, we obtain supersymmetric $AdS_2$ solutions from M2-branes wrapped on a topological disk. We consider four-dimensional $U(1)^4$-gauged $\mathcal{N}\,=\,2$ supergravity, \cite{Duff:1999gh}, where $U(1)^4$ is the maximal Abelian subgroup of $SO(8)_R$ symmetry in four dimensions. We study three subtruncations of the theory with different isometries: $SO(8)\rightarrow{S}O(2)\times{S}O(6)$, $SO(4)\times{S}O(4)$, and $SO(6)\times{S}O(2)$. In each subtruncation, we obtain supersymmetric $AdS_2$ solutions from M2-branes wrapped on a topological disk in Calabi-Yau two-, three-, and four-folds, respectively. Our solutions are natural generalizations of M2-branes wrapped on a constant curvature Riemann surface in \cite{Gauntlett:2001qs, Kim:2013xza}. Then we uplift the solutions to eleven-dimensional supergravity, \cite{Cremmer:1978km}, and calculate the Bekenstein-Hawking entropy.  For the solutions from topological disk in Calabi-Yau four-folds, the Bekenstein-Hawking entropy is finite and well-defined. On the other hand, from the topological disk in Calabi-Yau two- and three-folds, we could not find solutions with finite Bekenstein-Hawking entropy. It is parallel to the result of \cite{Gauntlett:2001qs}.

In section 2, we review $U(1)^4$-gauged $\mathcal{N}\,=\,2$ supergravity in four dimensions. In section 3, 4, and 5, we construct supersymmetric $AdS_2$ solutions from M2-branes wrapped on a topological disk in Calabi-Yau four-, two-, and three-folds, respectively. We conclude in section 6. The equations of motion are relegated in an appendix.

\bigskip

\noindent {\bf Note added:} After this work was posted on arXiv, \cite{Couzens:2021rlk} appeared which studies the spindle and topological disk solutions from M2-branes.

\bigskip

\noindent {\bf Note added 2:} Section 3.5 on flux quantization is added after publication as an addendum. We noticed that the result of the caluculation in this subsection was previously presented in (3.60)-(3.63) in section 3.2.2 of \cite{Couzens:2021rlk}. We present the calculation in detail in our conventions.

\section{$U(1)^4$-gauged $\mathcal{N}\,=\,2$ supergravity}

We review gauged $\mathcal{N}\,=\,2$ supergravity coupled to three vector multiplets in four dimensions, \cite{Duff:1999gh}. The bosonic field content is the metric and graviphoton from gravity multiplet and three scalar fields and three Abelian gauge fields from three vector multiplets. The Lagrangian is given by
\begin{equation}
\mathcal{L}\,=\,R-\frac{1}{2}\left(\partial\vec{\phi}\right)^2-2\sum_{\alpha=1}^{4}e^{\vec{a}_\alpha\cdot\vec{\phi}}F_\alpha^2-V\,,
\end{equation}
where $F_{\alpha,\mu\nu}\,=\,\partial_\mu{A}_{\alpha,\nu}-\partial_\nu{A}_{\alpha,\mu}$ and the scalar potential is
\begin{equation}
V\,=\,-4g^2\left(\cosh\phi_{12}+\cosh\phi_{13}+\cosh\phi_{14}\right)\,.
\end{equation}
We introduced the vectors of
\begin{equation}
\vec{\phi}\,=\,\left(\phi_{12},\,\phi_{13},\,\phi_{14}\right)\,,
\end{equation}
and
\begin{equation}
\vec{a}_1\,=\left(1,\,1,\,1\right)\,, \qquad \vec{a}_2\,=\left(1,\,-1,\,-1\right)\,, \qquad \vec{a}_3\,=\left(-1,\,1,\,-1\right)\,, \qquad \vec{a}_4\,=\left(-1,\,-1,\,1\right)\,.
\end{equation}
The equations of motion are presented in appendix A. 

The supersymmetry variations of spin 3/2- and 1/2-fields are given by
\begin{align}
\delta\psi_\mu\,^{\alpha{i}}\,=&\,\nabla_\mu\epsilon^{\alpha{i}}-2g\sum_{\beta=1}^4\Omega_{\alpha\beta}A_{\beta,\mu}\epsilon^{ij}\epsilon^{\alpha{j}}+\frac{g}{4\sqrt{2}}\sum_{\beta=1}^4e^{-\vec{a}_\beta\cdot\vec{\phi}/2}\gamma_\mu\epsilon^{\alpha{i}} \notag \\
&+\frac{1}{2\sqrt{2}}\sum_{\beta=1}^4\Omega_{\alpha\beta}e^{\vec{a}_\beta\cdot\vec{\phi}/2}F_{\beta,\nu\lambda}\gamma^{\nu\lambda}\gamma_{\mu}\epsilon^{ij}\epsilon^{\alpha{j}}\,, \notag \\
\delta\chi^{\alpha\,\beta{i}}\,=&\,-\frac{1}{\sqrt{2}}\gamma^\mu\partial_\mu\phi_{\alpha\beta}\epsilon^{ij}\epsilon^{\beta{j}}-g\sum_{\gamma,\delta=1}^4\Sigma_{\alpha\beta\gamma}\Omega_{\gamma\delta}e^{-\vec{a}_\delta\cdot\vec{\phi}/2}\epsilon^{ij}\epsilon^{\beta{j}} \notag \\
&+\sum_{\delta=1}^4\Omega_{\alpha\delta}e^{\vec{a}_\delta\cdot\vec{\phi}/2}F_{\delta,\mu\nu}\gamma^{\mu\nu}\epsilon^{\beta{i}}\,,
\end{align}
where we define the matrix,
\begin{equation}
\Omega\,=\,\frac{1}{2}\left(
\begin{array}{llll}
 1 & \,\,\,\, 1 & \,\,\,\, 1 & \,\,\,\,1 \\
 1 & \,\,\,\, 1 & -1 & -1 \\
 1 & -1 & \,\,\,\, 1 & -1 \\
 1 & -1 & -1 & \,\,\,\,1
\end{array}
\right)\,.
\end{equation}
and the tensor, $\Sigma_{\alpha\beta\gamma}$, selects a particular $\gamma$ depending on $\alpha$ and $\beta$,
\begin{equation}
\Sigma_{\alpha\beta\gamma}\,=\,\left\{ \begin{aligned}
& |\epsilon_{\alpha\beta\gamma}|\,, \qquad \alpha,\,\beta,\,\gamma\,\ne\,1\,, \\
& \delta_{\beta\gamma}\,, \qquad \,\,\,\,\,\,\, \alpha\,=\,1\,, \\
& \delta_{\alpha\gamma}\,, \qquad \,\,\,\,\,\,\, \beta\,=\,1\,, \\
& 0\,, \qquad \qquad \text{otherwise}\,.
\end{aligned}
\right.
\end{equation}

\section{Topological disk in Calabi-Yau four-folds}

\subsection{Supersymmetry equations}

We consider the background,
\begin{equation}
ds^2\,=\,f(y)ds_{AdS_2}^2+g_1(y)dy^2+g_2(y)dz^2\,,
\end{equation}
with the gauge fields,
\begin{equation}
A_1\,=\,A_2\,=A_3\,=\,A_z(y)dz\,, \qquad A_4\,=\,0\,,
\end{equation}
and the scalar fields,
\begin{equation}
\phi_{12}\,=\,\phi_{13}\,=\,-\phi_{14}\,=\,\phi(y)\,.
\end{equation}
The gamma matrices are given by
\begin{equation}
\gamma^{\alpha}\,=\,\rho^{\alpha}\otimes\sigma^2\,, \qquad \gamma^{\hat{y}}\,=\,1\otimes\sigma^1\,, \qquad \gamma^{\hat{z}}\,=\,1\otimes\sigma^3\,,
\end{equation}
where $\alpha$ are two-dimensional flat indices and the hatted indices are flat indiced for the corresponding coordinates. $\rho^\alpha$ are two-dimensional gamma matrices with $\{\rho^\alpha,\rho^\beta\}\,=\,2\eta^{\alpha\beta}$ and $\sigma^{1,2,3}$ are the Pauli matrices. The spinor is given by
\begin{equation}
\epsilon\,=\,\vartheta\otimes\eta\,,
\end{equation}
where $\vartheta$ is a Killing spinor on $AdS_2$ and $\eta\,=\,\eta(y,z)$. The Killing spinors satisfy
\begin{equation}
\nabla_\alpha^{AdS_2}\vartheta\,=\,\frac{1}{2}s\rho_\alpha\rho^*\vartheta\,,
\end{equation}
where $s\,=\,\pm1$ and $\rho^*$ is the two-dimensional chirality matrix.

The supersymmetry equations are obtained by setting the supersymmetry variations of the
fermionic fields to zero. From the supersymmetry variations, we obtain
\begin{align}
0\,=\,&-\frac{i}{2}s\gamma^{\hat{t}}\epsilon^{\alpha{i}}+\frac{1}{2}f^{1/2}g_1^{-1/2}\frac{1}{2}\frac{f'}{f}\gamma^{\hat{y}}\epsilon^{\alpha{i}}+\frac{g}{4\sqrt{2}}\left(3e^{-\frac{\phi}{2}}+e^{\frac{3\phi}{2}}\right)f^{1/2}\epsilon^{\alpha{i}} \notag \\
&+\frac{3}{2\sqrt{2}}e^{\frac{\phi}{2}}A'_z{g}_1^{-1/2}g_2^{-1/2}f^{1/2}\gamma^{\hat{y}\hat{z}}\epsilon^{ij}\epsilon^{\alpha{j}}\,, \notag \\
0\,=\,&\partial_y\epsilon^{\alpha{i}}+\frac{g}{4\sqrt{2}}\left(3e^{-\frac{\phi}{2}}+e^{\frac{3\phi}{2}}\right)g_1^{1/2}\gamma^{\hat{y}}\epsilon^{\alpha{i}}-\frac{3}{2\sqrt{2}}e^{\frac{\phi}{2}}A'_z{g}_2^{-1/2}\gamma^{\hat{z}}\epsilon^{ij}\epsilon^{\alpha{j}}\,, \notag \\
0\,=\,&\partial_z\epsilon^{\alpha{i}}-3gA_z\epsilon^{ij}\epsilon^{\alpha{j}}-\frac{1}{2}g_1^{-1/2}g_2^{1/2}\frac{1}{2}\frac{g_2'}{g_2}\gamma^{\hat{y}\hat{z}}\epsilon^{\alpha{i}}+\frac{g}{4\sqrt{2}}\left(3e^{-\frac{\phi}{2}}+e^{\frac{3\phi}{2}}\right)g_2^{1/2}\gamma^{\hat{z}}\epsilon^{\alpha{i}} \notag \\
&+\frac{3}{2\sqrt{2}}e^{\frac{\phi}{2}}A'_z{g}_1^{-1/2}\gamma^{\hat{y}}\epsilon^{ij}\epsilon^{\alpha{j}}\,, \notag \\
0\,=\,&-\frac{1}{4}g_1^{-1/2}\phi'\gamma^{\hat{y}}\epsilon^{ij}\epsilon^{\alpha{j}}-\frac{g}{4\sqrt{2}}\left(e^{-\frac{\phi}{2}}-e^{\frac{3\phi}{2}}\right)\epsilon^{ij}\epsilon^{\alpha{j}}+\frac{1}{2\sqrt{2}}e^{\frac{\phi}{2}}A'_z{g}_1^{-1/2}g_2^{-1/2}\gamma^{\hat{y}\hat{z}}\epsilon^{\alpha{i}}\,,
\end{align}
where the first three and the last equations are from the spin-3/2 and spin-1/2 field variations, respectively. By multiplying suitable functions and gamma matrices and adding the last equation to the first three equations, we obtain
\begin{align} \label{presusy}
0\,=\,&-\frac{i}{2}s\gamma^{\hat{t}}\epsilon^1+\frac{1}{2}f^{1/2}g_1^{-1/2}\frac{1}{2}\frac{f'}{f}\gamma^{\hat{y}}\epsilon^1-\frac{3}{4}f^{1/2}g_1^{-1/2}\phi'\gamma^{\hat{y}}\epsilon^1+\frac{g}{\sqrt{2}}e^{\frac{3\phi}{2}}f^{1/2}\epsilon^1\,, \notag \\
0\,=\,&\partial_y\epsilon^1-\frac{3}{4}\phi'\epsilon^1+\frac{g}{\sqrt{2}}e^{\frac{3\phi}{2}}g_1^{1/2}\gamma^{\hat{y}}\epsilon^1-\frac{3}{\sqrt{2}}e^{\frac{\phi}{2}}A'_z{g}_2^{-1/2}\gamma^{\hat{z}}\epsilon^2\,, \notag \\
0\,=\,&\partial_z\epsilon^1-3gA_z\epsilon^2+\frac{3}{\sqrt{2}}e^{\frac{\phi}{2}}A'_z{g}_1^{-1/2}\gamma^{\hat{y}}\epsilon^2+\frac{g}{\sqrt{2}}e^{\frac{3\phi}{2}}g_2^{1/2}\gamma^{\hat{z}}\epsilon^1-\frac{1}{2}g_1^{-1/2}g_2^{1/2}\frac{1}{2}\frac{g'_2}{g_2}\gamma^{\hat{y}\hat{z}}\epsilon^1+\frac{3}{4}g_1^{-1/2}g_2^{1/2}\phi'\gamma^{\hat{y}\hat{z}}\epsilon^1\,, \notag \\
0\,=\,&\frac{g}{2\sqrt{2}}\left(e^{-\frac{\phi}{2}}-e^{\frac{3\phi}{2}}\right)\epsilon^1+\frac{1}{2}g_1^{-1/2}\phi'\gamma^{\hat{y}}\epsilon^1+\frac{1}{\sqrt{2}}e^{\frac{\phi}{2}}A'_z{g}_1^{-1/2}g_2^{-1/2}\gamma^{\hat{y}\hat{z}}\epsilon^2\,.
\end{align}
The spinor is supposed to have a charge under the $U(1)_z$ isometry,
\begin{equation}
\eta(y,z)\,=\,e^{inz}\widehat{\eta}(y)\,,
\end{equation}
where $n$ is a constant. It shows up in the supersymmetry equations in the form of $\left(-i\partial_z+\frac{1}{2}A_z\right)\eta\,=\,\left(n+\frac{1}{2}A_z\right)\eta$ which is invariant under
\begin{equation}
A_\alpha\,\mapsto\,A_\alpha-2\alpha_0dz\,, \qquad \eta\,\mapsto\,e^{i\alpha_0z}\eta\,,
\end{equation}
where $\alpha_0$ is a constant. We also define
\begin{equation} \label{AAhat}
\frac{1}{2}\widehat{A}_z\,=\,n+\frac{1}{2}A_z\,.
\end{equation}
We solve the equation of motion for the gauge fields and obtain
\begin{equation} \label{solgaugep}
A_z'\,=\,be^{-\phi}g_1^{1/2}g_2^{1/2}f^{-1}\,,
\end{equation}
where $b$ is a constant. Employing the expressions we discussed in \eqref{presusy} beside the second equation, we finally obtain the supersymmetry equations,
\begin{align}
0\,=\,&-isf^{-1/2}\left(\sigma^3\eta\right)+\frac{1}{2}g_1^{-1/2}\left[\frac{f'}{f}-3\phi'\right]\left(\sigma^1\eta\right)+\sqrt{2}ge^{\frac{3\phi}{2}}\eta\,, \notag \\
0\,=\,&-3g\widehat{A}_z{g}_2^{-1/2}\left(\sigma^1\eta\right)+\frac{3}{\sqrt{2}}f^{-1}e^{-\frac{\phi}{2}}b\eta-\frac{1}{4}g_1^{-1/2}\left[\frac{g_2'}{g_2}-3\phi'\right]\left(\sigma^3\eta\right)-\frac{g}{\sqrt{2}}e^{\frac{3\phi}{2}}\left(i\sigma^2\eta\right)\,, \notag \\
0\,=\,&\frac{g}{2\sqrt{2}}\left(e^{-\frac{\phi}{2}}-e^{\frac{3\phi}{2}}\right)\eta+\frac{1}{2}g_1^{-1/2}\phi'\left(\sigma^1\eta\right)-\frac{1}{\sqrt{2}}f^{-1}e^{-\frac{\phi}{2}}b\left(i\sigma^2\eta\right)\,.
\end{align}

The supersymmetry equations are in the form of $M^{(i)}\eta\,=\,0$, $i\,=\,1,\,2,\,3$, where $M^{(i)}$ are three $2\times{2}$ matrices, as we follow \cite{Bah:2021mzw, Bah:2021hei},
\begin{equation}
M^{(i)}\,=\,X_0^{(i)}\mathbb{I}_2+X_1^{(i)}\sigma^1+X_2^{(i)}\left(i\sigma^2\right)+X_3^{(i)}\sigma^3\,.
\end{equation}
We rearrange the matrices to introduce $2\times{2}$ matrices,
\begin{equation}
\mathcal{A}^{ij}\,=\,\text{det}\left(v^{(i)}|w^{(j)}\right)\,, \qquad \mathcal{B}^{ij}\,=\,\text{det}\left(v^{(i)}|v^{(j)}\right)\,, \qquad \mathcal{C}^{ij}\,=\,\text{det}\left(w^{(i)}|w^{(j)}\right)\,,
\end{equation}
from the column vectors of
\begin{equation}
v^{(i)}\,=\,\left(
\begin{array}{l}
 X_1^{(i)}+X_2^{(i)} \\
 -X_0^{(i)}-X_3^{(i)}
\end{array}
\right)\,, \qquad
w^{(i)}\,=\,\left(
\begin{array}{l}
 X_0^{(i)}-X_3^{(i)} \\
 -X_1^{(i)}+X_2^{(i)}
\end{array}
\right)\,.
\end{equation}
We present the components of $X^{(i)}_a$, $a\,=\,0,\,1,\,2,\,3$,
\begin{align}
X^{(1)}_0\,=&\,0\,, \qquad \qquad \qquad \qquad \,\, X^{(2)}_0\,=\,i\frac{3}{\sqrt{2}}f^{-1}e^{-\frac{\phi}{2}}b\,, \qquad \qquad \qquad \,\, X^{3}_0\,=\,\frac{g}{2\sqrt{2}}\left(e^{-\frac{\phi}{2}}-e^{\frac{3\phi}{2}}\right)\,, \notag \\
X^{(1)}_1\,=&\,-isf^{-1/2}\,, \qquad \qquad \,\,\, X^{(2)}_1\,=\,-i3gg_2^{-1/2}\widehat{A}_z\,, \qquad \qquad \qquad \,\, X^{(3)}_1\,=\,\frac{1}{2}g_1^{-1/2}\phi'\,, \notag \\
X^{(1)}_2\,=&\,\frac{1}{2}g_1^{-1/2}\left[\frac{f'}{f}-3\phi'\right]\,, \,\,\,\,\,\, X^{(2)}_2\,=\,-\frac{g}{\sqrt{2}}e^{\frac{3\phi}{2}}\,, \qquad \qquad \qquad \qquad X^{(3)}_2\,=\,-i\frac{1}{\sqrt{2}}f^{-1}e^{-\frac{\phi}{2}}b\,, \notag \\
X^{(1)}_3\,=&\,\sqrt{2}ge^{\frac{3\phi}{2}}\,, \qquad \qquad \,\,\,\,\,\,\,\,\,\,\, X^{(2)}_3\,=\,-\frac{1}{4}g_1^{-1/2}\left[\frac{g_2'}{g_2}-3\phi'\right]\,, \qquad \,\,\,\,\, X^{(3)}_3\,=\,0\,.
\end{align}

From the vanishing of $\mathcal{A}^{ij}$, $\mathcal{B}^{ij}$ and $\mathcal{C}^{ij}$, necessary conditions for non-trivial solutions are obtained. From $\mathcal{A}^{ii}\,=\,0$, we find
\begin{align} \label{Adiag}
0\,=\,&\frac{1}{f}+\frac{1}{4g_1}\left(\frac{f'}{f}-3\phi'\right)^2-2g^2e^{3\phi}\,, \notag \\
0\,=\,&-\frac{9b^2e^{-\phi}}{2f^2}-\frac{1}{16g_1}\left(\frac{g_2'}{g_2}-3\phi'\right)^2+\frac{g^2}{2}e^{3\phi}+\frac{9g^2\widehat{A}_z^2}{g_2}\,, \notag \\
0\,=\,&-\frac{\left(\phi'\right)^2}{4g_1}-\frac{b^2e^{-\phi}}{2f^2}+\frac{g^2}{8}\left(e^{-\frac{\phi}{2}}-e^{\frac{3\phi}{2}}\right)^2\,.
\end{align}
From $\mathcal{A}^{ij}+\mathcal{A}^{ji}\,=\,0$, we find
\begin{align} \label{Aplus}
0\,=\,&-\frac{ge^{\frac{3\phi}{2}}}{\sqrt{2}\sqrt{g_1}}\left(\frac{f'}{f}-\frac{g_2'}{g_2}\right)+\frac{6sg\widehat{A}_z}{\sqrt{f}\sqrt{g_2}}\,, \notag \\
0\,=\,&-\frac{ibe^{-\frac{\phi}{2}}}{\sqrt{2}f\sqrt{g_1}}\left(\frac{f'}{f}-3\phi'\right)+\frac{is\phi'}{\sqrt{f}\sqrt{g_1}}\,, \notag \\
0\,=\,&\frac{igbe^{-\phi}}{f}+\frac{igb}{2f}e^{-\frac{\phi}{2}}\left(e^{-\frac{\phi}{2}}-e^{\frac{3\phi}{2}}\right)+\frac{i3g\widehat{A}_z\phi'}{\sqrt{g_1}\sqrt{g_2}}\,.
\end{align}
From $\mathcal{A}^{ij}-\mathcal{A}^{ji}\,=\,0$, we find
\begin{align} \label{Aminus}
0\,=\,&\frac{i6gbe^\phi}{f}+\frac{i\sqrt{2}sge^{\frac{3\phi}{2}}}{\sqrt{f}}+\frac{i3g\widehat{A}_z}{\sqrt{g_1}\sqrt{g_2}}\left(\frac{f'}{f}-3\phi'\right)\,, \notag \\
0\,=\,&\frac{\sqrt{2}sbe^{-\frac{\phi}{2}}}{f^{3/2}}+\frac{1}{2g_1}\phi'\left(\frac{f'}{f}-3\phi'\right)-g^2e^{\frac{3\phi}{2}}\left(e^{-\frac{\phi}{2}}-e^{\frac{3\phi}{2}}\right)\,, \notag \\
0\,=\,&-\frac{g}{4\sqrt{2}\sqrt{g_1}}\left(\frac{g_2'}{g_2}-3\phi'\right)\left(e^{-\frac{\phi}{2}}-e^{\frac{3\phi}{2}}\right)+\frac{ge^{\frac{3\phi}{2}}\phi'}{\sqrt{2}\sqrt{g_1}}-\frac{3\sqrt{2}gbe^{-\frac{\phi}{2}}\widehat{A}_z}{f\sqrt{g_2}}\,.
\end{align}
From $\mathcal{B}^{ij}+\mathcal{C}^{ij}\,=\,0$, we find 
\begin{align}
0\,=\,&-\frac{i3be^{-\frac{\phi}{2}}}{\sqrt{2}f\sqrt{g_1}}\left(\frac{f'}{f}-3\phi'\right)-\frac{is}{2\sqrt{f}\sqrt{g_1}}\left(\frac{g_2'}{g_2}-3\phi'\right)-\frac{i12g^2e^{\frac{3\phi}{2}}\widehat{A}_z}{\sqrt{2}\sqrt{g_2}}\,, \notag \\
0\,=\,&\frac{g}{2\sqrt{2}\sqrt{g_1}}\left(\frac{f'}{f}-3\phi'\right)\left(e^{-\frac{\phi}{2}}-e^{\frac{3\phi}{2}}\right)-\frac{2ge^{\frac{3\phi}{2}}\phi'}{\sqrt{2}\sqrt{g_1}}\,, \notag \\
0\,=\,&\frac{3b^2e^{-\phi}}{f^2}-\frac{1}{4g_1}\phi'\left(\frac{g_2'}{g_2}-3\phi'\right)+\frac{g^2}{2}e^{\frac{3\phi}{2}}\left(e^{-\frac{\phi}{2}}-e^{\frac{3\phi}{2}}\right)\,.
\end{align}
From $\mathcal{B}^{ij}-\mathcal{C}^{ij}\,=\,0$, we find 
\begin{align} \label{Abc}
0\,=\,&\frac{1}{4g_1}\left(\frac{f'}{f}-3\phi'\right)\left(\frac{g_2'}{g_2}-3\phi'\right)-\frac{3\sqrt{2}sbe^{-\frac{\phi}{2}}}{f^{3/2}}-2g^2e^{3\phi}\,, \notag \\
0\,=\,&\frac{i2gbe^\phi}{f}-\frac{isg}{\sqrt{2}\sqrt{f}}\left(e^{-\frac{\phi}{2}}-e^{\frac{3\phi}{2}}\right)\,, \notag \\
0\,=\,&\frac{ibe^{-\frac{\phi}{2}}}{2\sqrt{2}f\sqrt{g_1}}\left(\frac{g_2'}{g_2}-3\phi'\right)+\frac{i3be^{-\frac{\phi}{2}}\phi'}{\sqrt{2}f\sqrt{g_1}}+\frac{i3g^2\widehat{A}_z}{\sqrt{2}\sqrt{g_1}}\left(e^{-\frac{\phi}{2}}-e^{\frac{3\phi}{2}}\right)\,.
\end{align}

\subsection{Supersymmetric solutions}

From the second equation of \eqref{Abc}, we obtain
\begin{equation} \label{fsol}
f\,=\,\frac{8b^2e^{2\phi}}{\left(e^{-\frac{\phi}{2}}-e^{\frac{3\phi}{2}}\right)^2}\,.
\end{equation}
Then, from the third equation of \eqref{Adiag} with \eqref{fsol}, we obtain
\begin{equation}
g_1\,=\,\frac{32b^2\left(\phi'\right)^2}{\left(e^{-\frac{\phi}{2}}-e^{\frac{3\phi}{2}}\right)^2\left(16g^2b^2-e^{-5\phi}\left(e^{-\frac{\phi}{2}}-e^{\frac{3\phi}{2}}\right)^2\right)}\,.
\end{equation}
From the third equation of \eqref{Aplus}, we find an expression for $\sqrt{g_1}\sqrt{g_2}$,
\begin{equation} \label{g1g21}
\sqrt{g_1}\sqrt{g_2}\,=\,-\frac{6f\widehat{A}_z\phi'}{be^{\frac{\phi}{2}}\left(3e^{-\frac{3\phi}{2}}-e^{\frac{\phi}{2}}\right)}\,.
\end{equation}
Also from \eqref{solgaugep}, we find another expression for $\sqrt{g_1}\sqrt{g_2}$,
\begin{equation} \label{g1g22}
\sqrt{g_1}\sqrt{g_2}\,=\,\frac{e^\phi{f}\widehat{A}'_z}{b}\,.
\end{equation}
Equating \eqref{g1g21} and \eqref{g1g22}, we find an ordinary differential equation for $\widehat{A}_z$ and it gives
\begin{equation}
\widehat{A}_z\,=\,\mathcal{C}e^{-\frac{\phi}{2}}\left(3e^{-\frac{3\phi}{2}}-e^{\frac{\phi}{2}}\right)\,,
\end{equation}
where $\mathcal{C}$ is a constant. From \eqref{AAhat}, we find
\begin{equation}
A_z\,=\,\mathcal{C}e^{-\frac{\phi}{2}}\left(3e^{-\frac{3\phi}{2}}-e^{\frac{\phi}{2}}\right)+n\,.
\end{equation}
Then, from \eqref{g1g21} or \eqref{g1g22}, we obtain 
\begin{equation}
g_2\,=\,\frac{72\mathcal{C}^2e^{2\phi}\left(16g^2b^2-e^{-5\phi}\left(e^{-\frac{\phi}{2}}-e^{\frac{3\phi}{2}}\right)^2\right)}{\left(e^{-\frac{\phi}{2}}-e^{\frac{3\phi}{2}}\right)^2}\,.
\end{equation}
Therefore, we have determined all functions in terms of the scalar field, $\phi(y)$, and its derivative. The solution satisfies all the supersymmetry equations in \eqref{Adiag} to \eqref{Abc} and the equations motion which we present in appendix A. In order to satisfy the supersymmetry equations, the parameters should be one of the two cases,
\begin{align} \label{splus}
s\,=\,+1\,, \qquad b\,<\,0\,, \qquad \mathcal{C}\,>\,0\,, \\
s\,=\,-1\,, \qquad b\,>\,0\,, \qquad \mathcal{C}\,<\,0\,.
\end{align}
We can determine the scalar field by fixing the ambiguity in reparametrization of $y$ due to the covariance of the supersymmetry equations,
\begin{equation}
\phi(y)\,=\,\log{y}\,,
\end{equation}
where $y\,>\,0$.

Finally, let us summarize the solution. The metric is given by
\begin{equation} \label{metmet}
ds^2\,=\,\frac{8b^2y^3}{\left(1-y^2\right)^2}\left[ds_{AdS_2}^2+\frac{4}{h(y)y^4}dy^2+\frac{9\mathcal{C}^2h(y)}{b^2}dz^2\right]\,,
\end{equation}
where we define
\begin{equation}
h(y)\,=\,16g^2b^2-y^{-6}\left(1-y^2\right)^2\,.
\end{equation}
The gauge field is given by
\begin{equation}
\widehat{A}_z\,=\,\mathcal{C}\left(\frac{3}{y^2}-1\right)\,.
\end{equation}
The metric can also be written as
\begin{equation}
ds^2\,=\,\frac{8b^2y^3}{\left(1-y^2\right)^2}ds_{AdS_2}^2+\frac{32b^2}{h(y)y\left(1-y^2\right)^2}dy^2+\frac{72\mathcal{C}^2y^3h(y)}{\left(1-y^2\right)^2}dz^2\,.
\end{equation}

Now we consider the range of $y$ for regular solutions, $i.e.$, the metric functions are positive definite and the scalar fields are real.  We find regular solutions when we have
\begin{equation} \label{regrange}
1<y_1<y<\infty\,,
\end{equation}
where $y_1$ is determined from $h(y_1)\,=\,0$,
\begin{align} \label{rone}
y_1\,=\,\frac{1}{12gb}&\left[1+\left(1-216g^2b^2+12\sqrt{324g^4b^4-3g^2b^2}\right)^{1/3}\right. \notag \\
&+\left.\frac{1}{\left(1-216g^2b^2+12\sqrt{324g^4b^4-3g^2b^2}\right)^{1/3}}\right]\,.
\end{align}
We plot a representative solution with $g\,=\,1/4$, $b\,=\,-0.1$, and $\mathcal{C}\,=\,1$ in Figure 1. The metric on the space spanned by $\Sigma(y,z)$ in \eqref{metmet} has a topology of disk with the origin at $y\,=\,y_1$ and the boundary at $y\,=\,\infty$.

\begin{figure}[t]
\begin{center}
\includegraphics[width=2.0in]{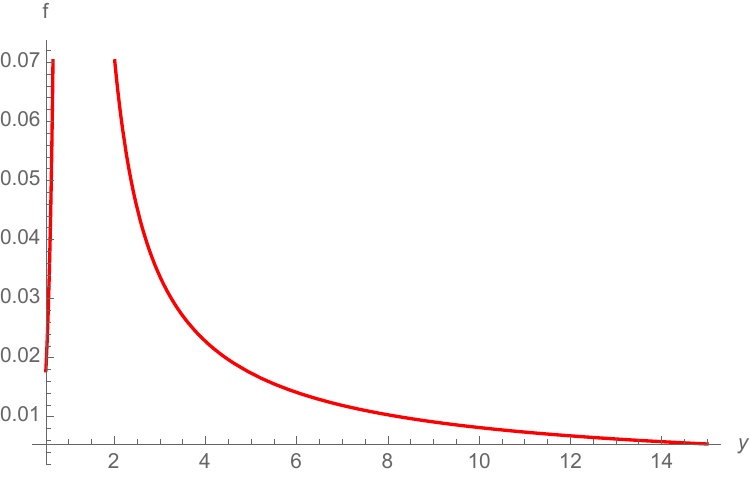} \qquad \includegraphics[width=2.0in]{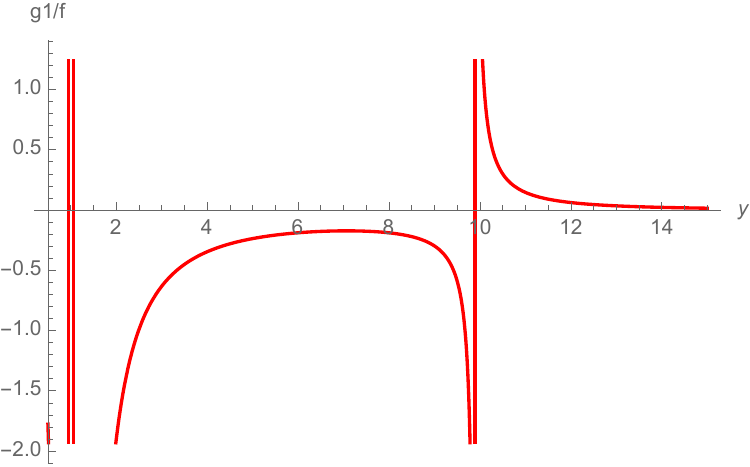} \qquad \includegraphics[width=2.0in]{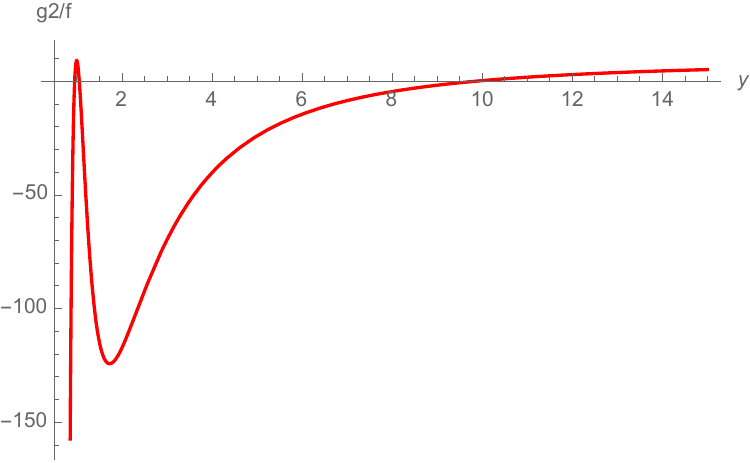}
\caption{{\it A representative solution with $g\,=\,1/4$, $b\,=\,-0.1$, and $\mathcal{C}\,=\,1$. The solution is regular in the range of $y_1=9.897\,<\,y\,<\,\infty$.}}
\end{center}
\end{figure}

Near $y\rightarrow\infty$ the $AdS_2$ warp factor vanishes and it is a curvature singularity of the metric,
\begin{equation} \label{wsingu}
ds^2\,\approx\,\frac{8b^2}{y}\left[ds_{AdS_2}^2+\frac{1}{4g^2b^2y^4}dy^2+144g^2\mathcal{C}^2dz^2\right]\,.
\end{equation}
This singularity is resolved when the solution is uplifted to eleven-dimensional supergravity.

Approaching $y\,=\,y_1$, the metric becomes to be 
\begin{equation}
ds^2\,=\,\frac{8b^2y_1^3}{\left(1-y_1^2\right)^2}\left[ds_{AdS_2}^2+\frac{16\Big[d\rho^2+\mathcal{C}^2\mathcal{E}^2(b)\rho^2dz^2\Big]}{-h'(y_1)r_1^4}\right]\,,
\end{equation}
where we introduced a new parametrization of coordinate, $\rho^2\,=\,y-y_1$ and $\mathcal{E}(b)$ is
\begin{equation} \label{defEb}
\mathcal{E}(b)\,=\,-\frac{3\left(y_1^4-4y_1^2+3\right)}{2by_1^5}\,.
\end{equation}
Then, the $\rho$-$z$ surface is locally an $\mathbb{R}^2/\mathbb{Z}_l$ orbifold if we set
\begin{equation} \label{cleb}
\mathcal{C}\,=\,\frac{1}{l\mathcal{E}(b)}\,,
\end{equation}
where $l\,=\,1,\,2,\,3,\ldots\,\,$.

Employing the Gauss-Bonnet theorem, we calculate the Euler characteristic of $\Sigma$, the $y$-$z$ surface, from \eqref{metmet}. The boundary at $y\,=\,\infty$ is a geodesic and thus has vanishing geodesic curvature. The only contribution to the Euler characteristic is  
\begin{equation}
\chi\left(\Sigma\right)\,=\,\frac{1}{4\pi}\int_\Sigma{R}_\Sigma\text{vol}_\Sigma\,=\,\frac{2\pi}{4\pi}\left(-\frac{3\mathcal{C}\left(y_1^4-4y_1^2+3\right)}{by_1^5}\right)\,=\,\mathcal{C}\mathcal{E}(b)\,=\,\frac{1}{l}\,,
\end{equation}
where $0\,<\,z\,<2\pi$. This result is natural for a disk in an $\mathbb{R}^2/\mathbb{Z}_l$ orbifold centered at $y\,=\,y_1$.

\subsection{Uplift to eleven-dimensional supergravity}

We uplift the solution on seven-sphere to eleven-dimensional supergravity, \cite{Cremmer:1978km}. where the seven-sphere gets warped and fibered. The bosonic fields in eleven-dimensional supergravity are the metric and the four-form flux. The uplift formula, \cite{Cvetic:1999xp}, is given for the metric,
\begin{equation}
ds_{11}^2\,=\,\Delta^{2/3}ds_4^2+\frac{2}{g^2}\frac{1}{\Delta^{1/3}}\sum_{I=1}^4\frac{1}{X_I}\left(d\mu_I^2+\mu_I^2\left(d\phi_I+2gA^I\right)^2\right)\,,
\end{equation}
and for the four-form flux,
\begin{align}
G_{(4)}\,=\,\sum_{I=1}^4&\left[\sqrt{2}gX_I\left(X_I\mu_I^2-\Delta\right)\text{vol}_4+\frac{1}{\sqrt{2}g}\frac{1}{X_I}d\left(\mu_I^2\right)\wedge*_4dX_I \right. \notag \\
&-\left.\frac{2\sqrt{2}}{g^2}\frac{1}{\left(X_I\right)^2}d\left(\mu_I^2\right)\wedge\left(d\phi_I+2gA^I\right)\wedge*_4F^I\right]\,,
\end{align}
where we define
\begin{equation}
\Delta\,=\,\sum_{I=1}^4X_I\mu_I^2\,,
\end{equation}
with the scalar fields,
\begin{equation}
X_I\,=\,e^{-\vec{a}_I\cdot\vec{\phi}/2}\,,
\end{equation}
and $\text{vol}_4$ and $*_4$ denote the volume form and the Hodge dual with respect to the four-dimensional metric, $ds_4^2$, respectively. The four-dimensional fields are $ds_5^2$, $F_I\,=\,dA_I$, and $X_I$. We introduce a parametrization in terms of angles on a three-sphere,
\begin{equation}
\mu_4\,=\,\sin\xi\,, \qquad \mu_1\,=\,\cos\xi\sin\varphi\,, \qquad \mu_2\,=\,\cos\xi\cos\varphi\sin\psi\,, \qquad \mu_3\,=\,\cos\xi\cos\varphi\cos\psi\,,
\end{equation}
and the ranges of the internal coordinates are
\begin{equation}
0\le\xi\,,\varphi\,,\psi\le\frac{\pi}{2}\,, \qquad 0\le\phi_1\,,\phi_2\,,\phi_3\,,\phi_4\le2\pi\,.
\end{equation}

By employing the uplift formula, we obtain the uplifted metric,
\begin{align} \label{upmet}
ds_{11}^2\,=\,\frac{8b^2y^3\Delta^{2/3}}{\left(1-y^2\right)^2}&\left[ds_{AdS_2}^2+\frac{4}{y^4h(y)}dy^2+\frac{9\mathcal{C}^2h(y)}{b^2}dz^2+\frac{\left(1-y^2\right)^2}{4g^2b^2y^4}d\xi^2\right. \notag \\ 
&+\frac{\left(1-y^2\right)^2}{4g^2b^2y^{5/2}\Delta}\cos^2\xi\Big(d\varphi^2+\sin^2\varphi{D}\phi_1^2+\cos^2\varphi\left(d\psi^2+\sin^2\psi{D}\phi_2^2+\cos^2\psi{D}\phi_3^2\right)\Big) \notag \\
&+\left.\frac{\left(1-y^2\right)^2}{4g^2b^2y^{9/2}\Delta}\sin^2\xi{d}\phi_4^2\right]\,.
\end{align}
where we have
\begin{equation}
\Delta\,=\,y^{3/2}\sin^2\xi+y^{-1/2}\cos^2\xi\,,
\end{equation}
and
\begin{equation}
D\phi_I\,=\,d\phi_I+2gA^I\,.
\end{equation}
We find the four-form flux to be
\begin{align}
G_{(4)}\,=\,&\frac{384\sqrt{2}gb^3\mathcal{C}y^4}{\left(1-y^2\right)^4}\left(y+2y^{-1/2}\Delta\right)dz\wedge{d}y\wedge\text{vol}_{AdS_2} \notag \\
+&\frac{12\sqrt{2}b\mathcal{C}y^4h}{g\left(1-y^2\right)^2}\sin(2\xi)\,dz\wedge{d}\xi\wedge\text{vol}_{AdS_2} \notag \\
-&\frac{4\sqrt{2}b}{g^2}\Big[\cos\xi\sin\varphi\left(\sin\xi\sin\varphi{d}\xi-\cos\xi\cos\varphi{d}\varphi\right)\wedge{D}\phi_1\wedge\text{vol}_{AdS_2} \notag \\
+&\cos\xi\cos\varphi\sin\psi\Big(\sin\xi\cos\varphi\sin\psi{d}\xi+\cos\xi\left(\sin\varphi\sin\psi{d}\varphi-\cos\varphi\cos\psi{d}\psi\right)\Big)\wedge{D}\phi_2\wedge\text{vol}_{AdS_2} \notag \\
+&\cos\xi\cos\varphi\cos\psi\Big(\sin\xi\cos\varphi\cos\psi{d}\xi+\cos\xi\left(\sin\varphi\cos\psi{d}\varphi+\cos\varphi\sin\psi{d}\psi\right)\Big)\wedge{D}\phi_3\wedge\text{vol}_{AdS_2}\Big]\,.
\end{align}
The Hodge dual of the four-form flux is
\begin{align}
*G_{(4)}\,=&\,\frac{16}{g^6\Delta^2}\left(y+2y^{-1/2}\Delta\right)\cos^5\xi\sin\xi\,d\xi\wedge\text{vol}_{S^5}\wedge\,d\phi_4 \notag \\
+&\frac{16}{g^6\Delta^2}\cos^6\xi\sin^2\xi\,dy\wedge\text{vol}_{S^5}\wedge\,d\phi_4 \notag \\
-&\frac{48\mathcal{C}}{g^5y^3}\cos^3\xi\sin\xi\cos^2\varphi\sin\left(2\psi\right) \notag \\
& \qquad \qquad \times\,d\xi\wedge\,d\psi\wedge\Big(\cos^2\varphi\,d\phi_2\wedge\,d\phi_3+\sin^2\varphi\,d\phi_1\wedge\left(d\phi_2-d\phi_3\right)\Big)\wedge\,d\phi_4\wedge\,dy\wedge\,dz \notag \\
+&\frac{24\mathcal{C}}{g^5y^3}\cos^3\xi\sin\xi\cos^3\varphi\csc\varphi\sin^2\left(2\psi\right)d\xi\wedge\,d\varphi\wedge\,d\phi_1\wedge\left(d\phi_2-d\phi_3\right)\wedge\,d\phi_4\wedge\,dy\wedge\,dz \notag \\
-&\frac{48\mathcal{C}}{g^5y^{3/2}\Delta}\cos^4\xi\sin^2\xi\cos^3\varphi\sin\varphi\sin\left(2\psi\right) \notag \\
& \qquad \,\,\,\,\,\,\,\,\,\, \times\,d\varphi\wedge\,d\psi\wedge\Big(d\phi_1\wedge\,d\phi_2+d\phi_2\wedge\,d\phi_3+d\phi_3\wedge\,d\phi_1\Big)\wedge\,d\phi_4\wedge\,dy\wedge\,dz\,,
\end{align}
where we define the volume form of gauged five-sphere,
\begin{equation}
\text{vol}_{S^5}\,=\,\cos^3\varphi\sin\varphi\cos\psi\sin\psi\,d\varphi\wedge\,d\psi\wedge\,D\phi_1\wedge\,D\phi_2\wedge\,D\phi_3\,.
\end{equation}

\subsection{Uplifted metric}

The nine-dimensional internal space of the uplifted metric is an $S_z^1\,\times\,S_{\phi_4}^1\,\times\,S^5$ fibration over the 2d base space, $B_2$, of $(y,\xi)$. The five-sphere, $S^5$, is spanned by $(\varphi,\psi,\phi_1,\phi_2,\phi_3)$. The 2d base space is a rectangle of $(y,\xi)$ over $[y_1,\infty)\,\times\left[0,\frac{\pi}{2}\right]$. See Figure 2. We explain the geometry of the internal space by three regions of the 2d base space, $B_2$.

\begin{itemize}
\item Region I: The side of $\mathsf{P}_1\mathsf{P}_2$.
\item Region II: The sides of $\mathsf{P}_2\mathsf{P}_3$ and $\mathsf{P}_3\mathsf{P}_4$.
\item Region III: The side of $\mathsf{P}_1\mathsf{P}_4$.
\end{itemize}

\begin{figure}[t]
\begin{center}
\includegraphics[width=4.5in]{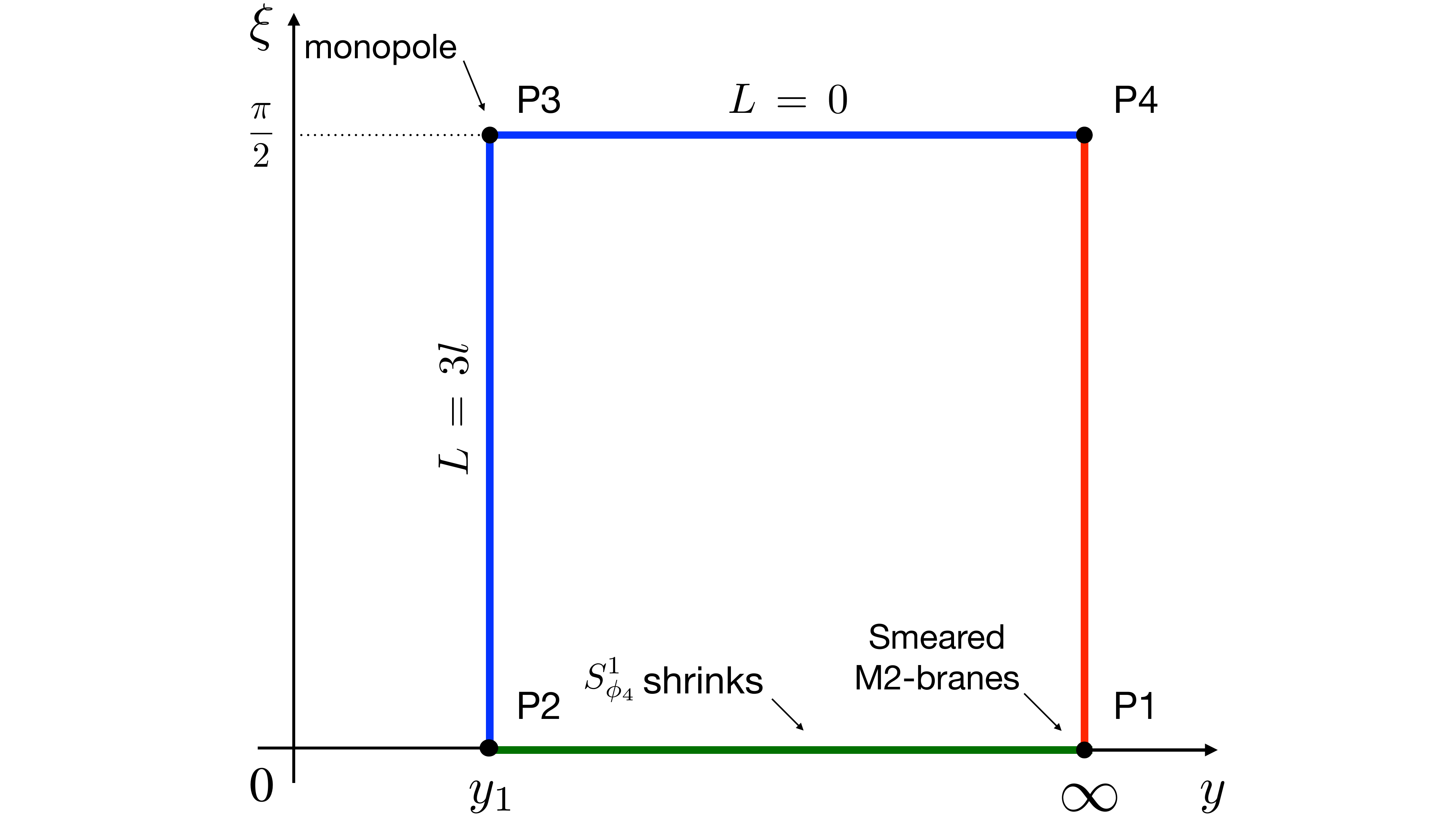}
\caption{{\it The two-dimensional base space, $B_2$, spanned by $y$ and $\xi$.}}
\end{center}
\end{figure}

\noindent {\bf Region I:} On the side of $\xi\,=\,0$, the circle, $S_{\phi_4}^1$, shrinks and the internal space caps off.

\bigskip

\noindent {\bf Region II: Monopole} We break $D\phi_1$, $D\phi_2$, and $D\phi_3$ and complete the square of $dz$, \cite{Bah:2021hei, Couzens:2021rlk}, to obtain the metric of
\begin{align}
ds_{11}^2\,=\,&\frac{8b^2y^3\Delta^{2/3}}{\left(1-y^2\right)^2}\left[ds_{AdS_2}^2+\frac{4}{y^4h}dy^2+\frac{\left(1-y^2\right)^2}{4g^2b^2y^4}d\xi^2+\frac{\left(1-y^2\right)^2}{4g^2b^2y^{5/2}\Delta}\cos^2\xi\left(d\varphi^2+\cos^2\varphi{d}\psi^2\right)\right. \notag \\
+&\frac{\left(1-y^2\right)^2}{4g^2b^2y^{9/2}\Delta}\sin^2\xi{d}\phi_4^2+R_z^2\left(dz+\frac{L}{\cos^2\xi}\left(\mu_1^2d\phi_1+\mu_2^2d\phi_2+\mu_3^2d\phi_3\right)\right)^2 \notag \\
+&\left.R_{\phi_1}^2\mu_1^2\Big(d\phi_1-L_1\left(\mu_2^2d\phi_2+\mu_3^2d\phi_3\right)\Big)^2+R_{\phi_2}^2\mu_2^2\Big(d\phi_2-L_2\mu_3^2d\phi_3\Big)^2+R_{\phi_3}^2\mu_3^2d\phi_3^2\right]\,.
\end{align}
The metric functions are defined to be
\begin{align}
R_{z}^2\,=&\,\frac{\mathcal{C}^2\left(9\Delta\,hy^{13/2}+\left(y^4-4y^2+3\right)^2\cos^2\xi\right)}{b^2y^{13/2}\Delta}\,, \notag \\
R_{\phi_1}^2\,=&\,\frac{\left(y^2-1\right)^2\left(9\Delta\,hy^{13/2}+\left(y^4-4y^2+3\right)^2\cos^2\xi\cos^2\varphi\right)}{4g^2b^2\Delta\,y^{5/2}\left(9\Delta\,hy^{13/2}+\left(y^4-4y^2+3\right)^2\cos^2\xi\right)}\,, \notag \\
R_{\phi_2}^2\,=&\,\frac{\left(y^2-1\right)^2\left(9\Delta\,hy^{13/2}+\left(y^4-4y^2+3\right)^2\cos^2\xi\cos^2\varphi\cos^2\psi\right)}{4g^2b^2\Delta\,y^{5/2}\left(9\Delta\,hy^{13/2}+\left(y^4-4y^2+3\right)^2\cos^2\xi\cos^2\varphi\right)}\,, \notag \\
R_{\phi_3}^2\,=&\,\frac{9hy^4\left(y^2-1\right)^2}{4g^2b^2\left(9\Delta\,hy^{13/2}+\left(y^4-4y^2+3\right)^2\cos^2\xi\cos^2\varphi\cos^2\psi\right)}\,,
\end{align}
with
\begin{align}
L\,=&\,-\frac{y^2\left(y^2-1\right)\left(y^4-4y^2+3\right)\cos^2\xi}{2g\mathcal{C}\left(9\Delta\,hy^{13/2}+\left(y^4-4y^2+3\right)^2\cos^2\xi\right)}\,, \notag \\
L_1\,=&\,\frac{\left(y^4-4y^2+3\right)^2}{9\Delta\,hy^{13/2}+\left(y^4-4y^2+3\right)^2\cos^2\xi\cos^2\varphi}\,, \notag \\
L_2\,=&\,\frac{\left(y^4-4y^2+3\right)^2}{9\Delta\,hy^{13/2}+\left(y^4-4y^2+3\right)^2\cos^2\xi\cos^2\varphi\cos^2\psi}\,.
\end{align}

The functions, $L(y,\xi)$, is piecewise constant along the sides of $y\,=\,y_1$ and $\xi\,=\,\frac{\pi}{2}$ of the 2d base, $B_2$,
\begin{equation}
L\left(y,\frac{\pi}{2}\right)\,=\,0\,, \qquad L\left(y_1,\xi\right)\,=\,\frac{3}{\mathcal{C}\mathcal{E}(b)}\,.
\end{equation}
The jump in $L$ at the corner, $(y,\xi)\,=\,\left(y_1,\frac{\pi}{2}\right)$, indicates the existence of a monopole source for the $Dz$ fibrations.

We perform a coordinate transformation of $\cos^2\xi\,=\,1-\mu^2$ and then $(y,\mu)$ to $(R,\Theta)$ defined by
\begin{equation}
\mu\,=\,1-\frac{1}{2}R^2\cos^2\frac{\Theta}{2}\,, \qquad y\,=\,y_1+\frac{y_1^4-4y_1^2+3}{4y_1}R^2\sin^2\frac{\Theta}{2}\,.
\end{equation}
In the limit of $R\,\rightarrow\,0$, the metric becomes
\begin{align}
&g^2ds_{11}^2\,\approx\,y_1^{-1/2}\left(\frac{1}{2}ds_{AdS_2}^2+d\phi_4^2\right) \notag \\
&+\,dR^2+R^2\left\{d\Theta^2+\cos^2\frac{\Theta}{2}\left(d\varphi^2+\cos^2\varphi{d}\psi^2+\sin^2\varphi{d}\phi_1^2+\cos^2\varphi\sin^2\psi{d}\phi_2^2+\cos^2\varphi\cos^2\psi{d}\phi_3^2\right)\right. \notag \\
&+\left.\left(5-4\cos\Theta\right)\frac{\mathcal{C}^2\mathcal{E}^2(b)}{9}\left[dz-\frac{1+\cos\Theta}{2\left(5-4\cos\Theta\right)}\frac{3}{\mathcal{C}\mathcal{E}(b)}\left(\sin^2\varphi{d}\phi_1+\cos^2\varphi\sin^2\psi{d}\phi_2+\cos^2\varphi\cos^2\psi{d}\phi_3\right)\right]^2\right\}\,.
\end{align}
The metric in curly bracket is $S^7/\mathbb{Z}_l$. When it is combined with the radial direction, R, we obtain the metric on $\mathbb{R}^8/\mathbb{Z}_l$. The geometry is overall $AdS_2\,\times\,S^1_{\phi_4}\,\times\,\mathbb{R}^8/\mathbb{Z}_l$. For $l>0$ there is an orbifold singularity at the location of the monopole and is smooth elsewhere.

\bigskip

\noindent {\bf Region III: Smeared M2-branes} The singularity at $y\rightarrow\infty$ in the warp factor of four-dimensional metric, \eqref{wsingu}, has been resolved in the uplifted metric, \eqref{upmet}. On the other hand, there is a singularity at $\left(y\rightarrow\infty,\,\sin\xi\rightarrow{0}\right)$ and we consider this singularity. We introduce coordinates, $(R,\Xi)$, for a reparametrization of $(y,\xi)$,
\begin{equation}
y\,=\,\frac{1}{R^{1/2}}\,, \qquad \sin\xi\,=\,R^{1/2}\cos\Xi\,.
\end{equation}
As $y\,\rightarrow\,\infty$ and $\sin\xi\,\rightarrow\,0$, or, equivalently, $R\,\rightarrow\,0$ and $\cos\Xi\,\rightarrow\,0$, the uplifted metric becomes
\begin{align}
ds_{11}^2\,\approx\,&8b^2R^{2/3}\cos^{4/3}\Xi\Big[ds_{AdS_2}^2+144g^2\mathcal{C}^2dz^2\Big] \notag \\
+&\frac{2}{g^2}\frac{1}{R^{1/3}}\cos^{4/3}\Xi\left[\frac{1}{4}dR^2+R^2\left(d\phi_4^2+\sin^2\Xi\,d\Xi^2\right) \right. \notag \\
&+\left.\frac{1}{\cos^2\Xi}\Big(d\varphi^2+\sin^2\varphi{D}\phi_1^2+\cos^2\varphi\left(d\psi^2+\sin^2\psi{D}\phi_2^2+\cos^2\psi{D}\phi_3^2\right)\Big)\right]\,.
\end{align}
The metric implies the smeared M2-brane sources. The M2-branes are 
\begin{itemize}
\item extended along the $AdS_2$ and $z$ directions;
\item localized at the origin of the $\mathbb{R}^3$ parametrized by $\phi_4$, $\Xi$ and $R$;
\item smeared along the five-sphere from $\left(\varphi,\,\psi,\,\phi_1,\,\phi_2,\,\phi_3\right)$ directions. 
\end{itemize}
The $R^{2/3}$ factor of the space where the M2-branes are extended and the $1/R^{1/3}$ factor of the space where the M2-branes are localized and smeared corresponds to the harmonic functions of $H^{2/3}$ and $H^{-1/3}$ of the black M2-branes, respectively.

Lastly, we briefly present the comparison of our geometry with the geometry of wrapped M5-branes in \cite{Bah:2021mzw, Bah:2021hei}. The overall geometries are given by 
\begin{align} \label{compar}
\text{Wrapped M2-branes}: \qquad &AdS_2\,\times\,S_{\phi_4}^1\,\times\,S_z^1\,\times\,S^5(\varphi,\psi,D\phi_1,D\phi_2,D\phi_3)\,\times\,[y,\xi]\,. \notag \\
\text{Wrapped M5-branes}: \qquad &AdS_5\,\times\,\,S^2\,\,\times\,\,S_z^1\,\times\,S_\phi^1(D\phi)\,\times\,[w,\mu]\,,
\end{align}
where we denote the gauged coordinates with $D$, $e.g.$, $D\phi$. For each metric, we presented the factors in the same order so that the corresponding factors are easily found. 

\subsection{Flux quantization}

We noticed that the result of the caluculation in this subsection was previously presented in (3.60)-(3.63) in section 3.2.2 of \cite{Couzens:2021rlk}. We present the calculation in detail in our conventions.

We consider the flux quantization conditions for the four-form flux. The integral of the four-form flux over any four-cycle in the internal space is an integer, $e.g.$, \cite{Ferrero:2020twa},
\begin{equation}
\frac{1}{\left(2\pi{l}_p\right)^6}\int*G_{(4)}\,\in\,\mathbb{Z}\,,
\end{equation}
where $l_p$ is the Planck length.

First, we consider the $*G_{(4)\xi\varphi\psi\phi_1\phi_2\phi_3\phi_4}$ component and we obtain
\begin{align} \label{m2N}
\frac{1}{\left(2\pi{l}_p\right)^6}\int*G_{(4)\xi\varphi\psi\phi_1\phi_2\phi_3\phi_4}\,=&\,\frac{1}{\left(2\pi{l}_p\right)^6}\int\frac{16}{g^6\Delta^2}\left(y+2y^{-1/2}\Delta\right)\cos^5\xi\sin\xi\,d\xi\wedge\text{vol}_{S^5}\wedge\,d\phi_4 \notag \\
=&\,\frac{1}{4\pi^2l_p^6g^6}\,\equiv\,N\,,
\end{align}
where $\text{vol}_{S^5}=\pi^3$ and $N\in\mathbb{N}$ is the number of M2-branes wrapping the two-dimensional manifold, $\Sigma$. This integration contour corresponds to the interval, $\mathsf{P}_1\mathsf{P}_2$, in figure 2.

Second, we consider the following three components and obtain
\begin{align}
&\frac{1}{\left(2\pi{l}_p\right)^6}\int*G_{(4)\xi\psi\phi_2\phi_3\phi_4yz} \notag \\
=&\,\frac{1}{\left(2\pi{l}_p\right)^6}\int\left(-\frac{48\mathcal{C}}{g^5y^3}\cos^3\xi\sin\xi\cos^2\varphi\sin\left(2\psi\right)\right)\cos^2\varphi\,d\xi\wedge\,d\psi\wedge\,d\phi_2\wedge\,d\phi_3\wedge\,d\phi_4\wedge\,dy\wedge\,dz \notag \\
=&\,-N\frac{6g\,\mathcal{C}}{y_1^2}\cos^2\varphi\cos^2\varphi\,,
\end{align}
\begin{align}
&\frac{1}{\left(2\pi{l}_p\right)^6}\int*G_{(4)\xi\psi\phi_1\phi_2\phi_4yz} \notag \\
=&\,\frac{1}{\left(2\pi{l}_p\right)^6}\int\left(-\frac{48\mathcal{C}}{g^5y^3}\cos^3\xi\sin\xi\cos^2\varphi\sin\left(2\psi\right)\right)\sin^2\varphi\,d\xi\wedge\,d\psi\wedge\,d\phi_1\wedge\,d\phi_2\wedge\,d\phi_4\wedge\,dy\wedge\,dz \notag \\
=&\,-N\frac{6g\,\mathcal{C}}{y_1^2}\cos^2\varphi\sin^2\varphi\,,
\end{align}
\begin{align}
&\frac{1}{\left(2\pi{l}_p\right)^6}\int*G_{(4)\xi\psi\phi_1\phi_3\phi_4yz} \notag \\
=&\,\frac{1}{\left(2\pi{l}_p\right)^6}\int\left(-\frac{48\mathcal{C}}{g^5y^3}\cos^3\xi\sin\xi\cos^2\varphi\sin\left(2\psi\right)\right)\left(-\sin^2\varphi\right)d\xi\wedge\,d\psi\wedge\,d\phi_1\wedge\,d\phi_3\wedge\,d\phi_4\wedge\,dy\wedge\,dz \notag \\
=&\,N\frac{6g\,\mathcal{C}}{y_1^2}\cos^2\varphi\sin^2\varphi\,.
\end{align}
In order to have identical results from these components, we fix $\psi=\frac{\pi}{4}$. Plugging $l_p$ from \eqref{m2N}, we obtain
\begin{align} \label{fluxk}
-\frac{1}{\left(2\pi{l}_p\right)^6}\int*G_{(4)\xi\psi\phi_2\phi_3\phi_4yz}\,=&\,-\frac{1}{\left(2\pi{l}_p\right)^6}\int*G_{(4)\xi\psi\phi_1\phi_2\phi_4yz}\,=\,\frac{1}{\left(2\pi{l}_p\right)^6}\int*G_{(4)\xi\psi\phi_1\phi_3\phi_4yz} \notag \\
=&\,N\frac{3g\,\mathcal{C}}{2y_1^2}\,\equiv\,K\,,
\end{align}
where $K\in\mathbb{N}$ is another integer.

From \eqref{defEb} and $y_1$ from \eqref{fluxk}, we find
\begin{equation} \label{defb1}
b\,=\,\frac{K^2\left(3g\left(N-2K\right)\right)}{6\sqrt{6}g\left(g\,\mathcal{C}NK\right)^{3/2}}\,,
\end{equation}
and also find
\begin{equation} \label{y1simple}
y_1^2\,=\,\frac{3g\,\mathcal{C}N}{2K}\,.
\end{equation}
Then, by plugging $y_1(b)$, \eqref{y1simple}, in \eqref{cleb} with the expression of $\mathcal{E}(b)$ in \eqref{defEb}, we also find another expression for $b$,
\begin{equation} \label{defb2}
b\,=\,-\frac{\mathcal{C}l\left(3g^2\mathcal{C}^2N^2-8g\,\mathcal{C}KN+4K^2\right)}{\sqrt{6}K^2}\left(\frac{K}{g\,\mathcal{C}N}\right)^{5/2}\,.
\end{equation}
Finally, identifying \eqref{defb1} and \eqref{defb2} we can solve for $\mathcal{C}$ and then for $b$ in terms of the quantum numbers, $N$ and $K$,
\begin{equation}
\mathcal{C}\,=\,\frac{12Kl+N}{6gNl}\,, \qquad b\,=\,\frac{\sqrt{Kl}\left(8Kl+N\right)}{2g\left(12Kl+N\right)^{3/2}}\,.
\end{equation}

\subsection{The Bekenstein-Hawking entropy}

Now we calculate the Bekenstein-Hawking entropy. For the metric of the form,
\begin{equation}
ds_{11}^2\,=\,e^{2\mathcal{A}}\left(ds_{AdS_2}^2+ds^2_{M_9}\right)\,,
\end{equation}
the Bekenstein-Hawking entropy is given by
\begin{equation} \label{bhformula}
S_{\text{BH}}\,=\,\frac{1}{G_N^{(11)}}\int_{M_9}e^{9\mathcal{A}}\text{vol}_{M_9}\,,
\end{equation}
where the eleven-dimensional gravitational constant is $G_N^{(11)}\,=\,\frac{\left(2\pi\right)^8l_p^9}{16\pi}$. Employing the formula, we obtain
\begin{align} \label{cformula}
S_{\text{BH}}\,=&\,\frac{16\sqrt{2}b\mathcal{C}}{\pi^2l_p^9g^7}\int_{y_\text{min}}^{y_\text{max}}\frac{y}{\left(1-y^2\right)^2}dy\,=\,\frac{8\sqrt{2}b\,\mathcal{C}}{\pi^2l_p^9g^7}\frac{1}{y_1^2-1}\,=\,\frac{64\sqrt{2}\pi}{3}\sqrt{\frac{K^3Nl}{12Kl+N}}\,,
\end{align}
where $y_{\text{min}}=y_1$ and $y_{\text{max}}=\infty$ for our solutions and $y_1=y_1(b)$ is given in \eqref{rone}. If we set $K\sim\,N$, the Bekenstein-Hawking entropy scales as $S_\text{BH}\sim\,N^{3/2}$ as the ABJM theory. Even though the uplifted solutions have singularities, we obtain a well-defined finite result for central charge. This result should match (3.63) in \cite{Couzens:2021rlk}.

\section{Solutions from Calabi-Yau two-folds}

\subsection{Supersymmetry equations}

As we proceed similar to the case of topological disk in Calabi-Yau four-folds, we will be brief. We consider the background,
\begin{equation}
ds^2\,=\,f(y)ds_{AdS_2}^2+g_1(y)dy^2+g_2(y)dz^2\,,
\end{equation}
with the gauge fields,
\begin{equation}
A_1\,=\,A_z(y)dz\,, \qquad A_2\,=\,A_3\,=\,A_4\,=\,0\,,
\end{equation}
and the scalar fields,
\begin{equation}
\phi_{12}\,=\,\phi_{13}\,=\,\phi_{14}\,=\,\phi(y)\,.
\end{equation}

We solve the equation of motion for the gauge fields and obtain
\begin{equation}
A_z'\,=\,b\,e^{-3\phi}g_1^{1/2}g_2^{1/2}f^{-1}\,.
\end{equation}
We present the supersymmetry equations,
\begin{align}
0\,=\,&-isf^{-1/2}\left(\sigma^3\eta\right)+\frac{1}{2}g_1^{-1/2}\left[\frac{f'}{f}-\phi'\right]\left(\sigma^1\eta\right)+\sqrt{2}ge^{\frac{\phi}{2}}\eta\,, \notag \\
0\,=\,&-g\widehat{A}_z{g}_2^{-1/2}\left(\sigma^1\eta\right)+\frac{1}{\sqrt{2}}f^{-1}e^{-\frac{3\phi}{2}}b\eta-\frac{1}{4}g_1^{-1/2}\left[\frac{g_2'}{g_2}-\phi'\right]\left(\sigma^3\eta\right)-\frac{g}{\sqrt{2}}e^{\frac{\phi}{2}}\left(i\sigma^2\eta\right)\,, \notag \\
0\,=\,&\frac{g}{2\sqrt{2}}\left(e^{-\frac{3\phi}{2}}-e^{\frac{\phi}{2}}\right)\eta+\frac{1}{2}g_1^{-1/2}\phi'\left(\sigma^1\eta\right)-\frac{1}{\sqrt{2}}f^{-1}e^{-\frac{3\phi}{2}}b\left(i\sigma^2\eta\right)\,,
\end{align}
where $s\,=\,\pm1$.

\subsection{Supersymmetric solutions}

We present the supersymmetric solutions,
\begin{align}
f\,=&\,\frac{8b^2}{e^{2\phi}\left(e^{-\frac{3\phi}{2}}-e^{\frac{\phi}{2}}\right)^2}\,, \notag \\
g_1\,=&\,\frac{32b^2\left(\phi'\right)^2}{\left(e^{-\frac{3\phi}{2}}-e^{\frac{\phi}{2}}\right)^2\left(16g^2b^2-e^\phi\left(e^{-\frac{3\phi}{2}}-e^{\frac{\phi}{2}}\right)^2\right)}\,, \notag \\
g_2\,=&\,\frac{8\mathcal{C}^2\left(16g^2b^2-e^\phi\left(e^{-\frac{3\phi}{2}}-e^{\frac{\phi}{2}}\right)^2\right)}{e^{2\phi}\left(e^{-\frac{3\phi}{2}}-e^{\frac{\phi}{2}}\right)^2}\,, \notag \\
A_z\,=&\,\mathcal{C}e^{-\frac{\phi}{2}}\left(e^{-\frac{3\phi}{2}}+e^{\frac{\phi}{2}}\right)\,,
\end{align}
where $\mathcal{C}$ is a constant. Therefore, we have determined all the functions in terms of the scalar field, $\phi(y)$, and its derivative. The solution satisfies all the supersymmetry equations and the equations motion which we present in appendix A. We can determine the scalar field by fixing the ambiguity in reparametrization of $y$ due to the covariance of the supersymmetry equations,
\begin{equation}
\phi(y)\,=\,\log{y}\,,
\end{equation}
where $y\,>\,0$.

Finally, let us summarize the solution. The metric is given by
\begin{equation} \label{metmet2}
ds^2\,=\,\frac{8b^2y}{\left(1-y^2\right)^2}\left[ds_{AdS_2}^2+\frac{4}{h(y)}dy^2+\frac{\mathcal{C}^2h(y)}{b^2}dz^2\right]\,
\end{equation}
where we define
\begin{equation}
h(y)\,=\,16g^2b^2-y^{-2}\left(1-y^2\right)^2\,.
\end{equation}
The gauge field is given by
\begin{equation}
\widehat{A}_z\,=\,\mathcal{C}\left(1+\frac{1}{y^2}\right)\,.
\end{equation}

For the choice of $s=+1$, \eqref{splus}, we find a class of solutions when we have
\begin{equation} \label{regrange2}
y_1<y<y_2\,,
\end{equation}
where $y_1$ and $y_2$ is obtained from $h(y)\,=\,0$,
\begin{equation} \label{rone2}
y_1\,=\,2gb+\sqrt{1+4g^2b^2}\,, \qquad y_2\,=\,-2gb+\sqrt{1+4g^2b^2}\,,
\end{equation}
and $y_1<1<y_2$. We plot a representative solution with $g=1/4$, $b=-1.5$ and $\mathcal{C}\,=\,1$ in Figure 3.

\begin{figure}[t]
\begin{center}
\includegraphics[width=2.0in]{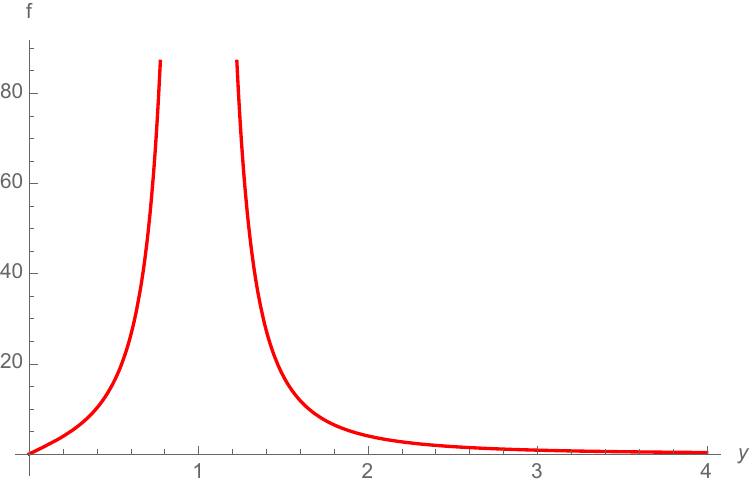} \qquad \includegraphics[width=2.0in]{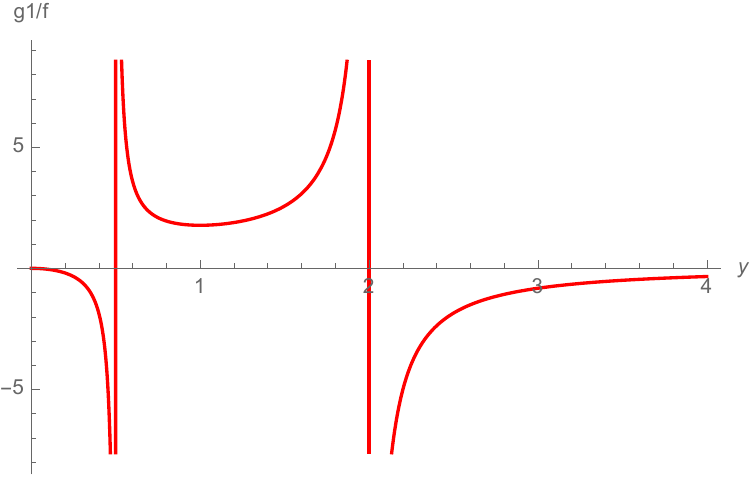} \qquad \includegraphics[width=2.0in]{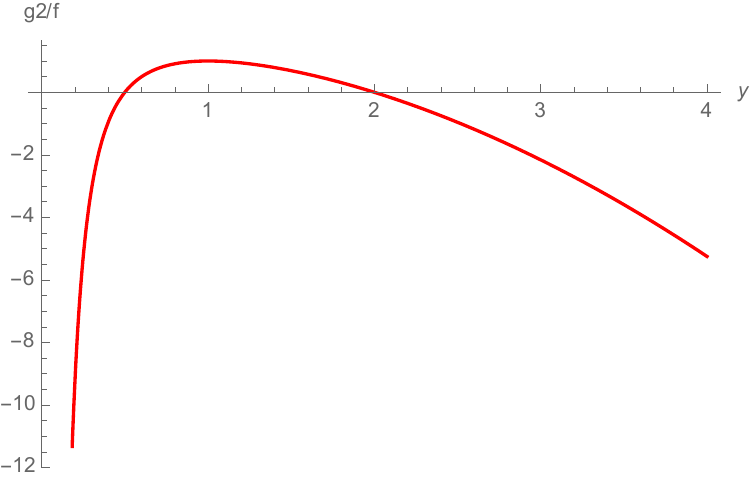}
\caption{{\it A representative solution with $g=1/4$, $b=-1.5$ and $\mathcal{C}\,=\,1$. The solution is regular in the range of $y_1\,=\,0.5\,\,<\,y\,<\,y_2\,=\,2.0$.}}
\end{center}
\end{figure}

Approaching $y\,=\,y_*$, where $y_*\,=\,y_1\,,y_2$, the metric becomes to be 
\begin{equation}
ds^2\,=\,\frac{8b^2y}{\left(1-y^2\right)^2}\left[ds_{AdS_2}^2+\frac{16\Big[d\rho^2+\mathcal{C}^2\mathcal{E}_*(b)^2\rho^2dz^2\Big]}{-h'(y_*)}\right]\,,
\end{equation}
where we introduced a new parametrization of coordinate, $\rho^2\,=\,|y_*-y|$ and $\mathcal{E}_*(b)$ are
\begin{equation}
\mathcal{E}_1(b)\,=\,\frac{h'(y_1)}{4b}\,, \qquad \mathcal{E}_2(b)\,=\,\frac{h'(y_2)}{4b}\,,
\end{equation}
for $y=y_1$ and $y=y_2$, respectively. Then, the $\rho$-$z$ surface is locally an $\mathbb{R}^2/\mathbb{Z}_l$ orbifold if we set
\begin{equation}
l_1\,=\,-\frac{1}{\mathcal{C}\,\mathcal{E}_1(b)}\,, \qquad l_2\,=\,\frac{1}{\mathcal{C}\,\mathcal{E}_2(b)}\,,
\end{equation}
where $l_1\,,l_2\,=\,1,\,2,\,3,\ldots\,\,$.

Employing the Gauss-Bonnet theorem, we calculate the Euler characteristic of $\Sigma$, the $y$-$z$ surface, from \eqref{metmet2}. The only contribution to the Euler characteristic is  
\begin{equation}
\chi\left(\Sigma\right)\,=\,\frac{1}{4\pi}\int_\Sigma{R}_\Sigma\text{vol}_\Sigma\,=\,8\,\mathcal{C}g\left(1+4g^2b^2\right)\,=\,\frac{1}{l_1}+\frac{1}{l_2}\,,
\end{equation}
where $0\,<\,z\,<2\pi$.

\subsection{The Bekenstein-Hawking entropy}

We uplift the solution to eleven-dimensional supergravity. The only non-trivial fields in eleven-dimensional supergravity are the metric and the four-form flux. We introduce a parametrization in terms of angles on a three-sphere,
\begin{equation}
\mu_1\,=\,\sin\xi\,, \qquad \mu_2\,=\,\cos\xi\sin\varphi\,, \qquad \mu_3\,=\,\cos\xi\cos\varphi\sin\psi\,, \qquad \mu_4\,=\,\cos\xi\cos\varphi\cos\psi\,.
\end{equation}
By employing the uplift formula, we obtain the uplifted metric,
\begin{align} \label{upmet2}
ds_{11}^2\,=\,\frac{8b^2y\Delta^{2/3}}{\left(1-y^2\right)^2}&\left[ds_{AdS_2}^2+\frac{4}{h(y)}dy^2+\frac{\mathcal{C}^2h(y)}{b^2}dz^2+\frac{\left(1-y^2\right)^2}{4g^2b^2}d\xi^2\right. \notag \\ 
&+\frac{\left(1-y^2\right)^2}{4g^2b^2y^{3/2}\Delta}\cos^2\xi\Big(d\varphi^2+\sin^2\varphi{D}\phi_2^2+\cos^2\varphi\left(d\psi^2+\sin^2\psi{D}\phi_3^2+\cos^2\psi{D}\phi_4^2\right)\Big) \notag \\
&+\left.\frac{y^{1/2}\left(1-y^2\right)^2}{4g^2b^2\Delta}\sin^2\xi{d}\phi_1^2\right]\,,
\end{align}
where we have
\begin{equation}
\Delta\,=\,y^{-3/2}\sin^2\xi+y^{1/2}\cos^2\xi\,.
\end{equation}

The Bekenstein-Hawking entropy is calculated by employing the formula, \eqref{bhformula},
\begin{align}
S_{\text{BH}}\,=&\,\frac{16\sqrt{2}b\mathcal{C}}{3\pi^2l_p^9g^7}\int_{y_\text{min}}^{y_\text{max}}\frac{y}{\left(1-y^2\right)^2}dy\,,
\end{align}
where $y_{\text{min}}=y_1$ and $y_{\text{max}}=y_2$ for our solutions and $y_1=y_1(b)$ and $y_2=y_2(b)$ are given in \eqref{rone2}. However, the integral diverges at $y\,=\,1$. 

Unlike the singularity of the warp factor, \eqref{wsingu}, which is resolved when uplifted, the singularity at $y=1$ is not resolved in the uplift. In order to avoid this singularity, we have to find solution which is well-defined in the range away from $y=1$. In the truncation we consider in this section, we were not able to find a solution of such range.

\section{Solutions from Calabi-Yau three-folds}

\subsection{Supersymmetry equations}

As we proceed similar to the case of topological disk in Calabi-Yau four-folds, we will be brief. We consider the background,
\begin{equation}
ds^2\,=\,f(y)ds_{AdS_2}^2+g_1(y)dy^2+g_2(y)dz^2\,,
\end{equation}
with the gauge fields,
\begin{equation}
A_1\,=\,A_2\,=\,A_z(y)dz\,, \qquad A_3\,=\,A_4\,=\,0\,,
\end{equation}
and the scalar fields,
\begin{equation}
\phi_{12}\,=\,\phi(y)\,, \qquad \phi_{13}\,=\,\phi_{14}\,=\,0\,.
\end{equation}

We solve the equation of motion for the gauge fields and obtain
\begin{equation}
A_z'\,=\,b\,e^{-\phi}g_1^{1/2}g_2^{1/2}f^{-1}\,,
\end{equation}
where $b$ is a constant. We present the supersymmetry equations,
\begin{align}
0\,=\,&-isf^{-1/2}\left(\sigma^3\eta\right)+\frac{1}{2}g_1^{-1/2}\left[\frac{f'}{f}-\phi'\right]\left(\sigma^1\eta\right)+\sqrt{2}ge^{\frac{\phi}{2}}\eta\,, \notag \\
0\,=\,&-g\widehat{A}_z{g}_2^{-1/2}\left(\sigma^1\eta\right)+\sqrt{2}f^{-1}e^{-\frac{\phi}{2}}b\eta-\frac{1}{4}g_1^{-1/2}\left[\frac{g_2'}{g_2}-\phi'\right]\left(\sigma^3\eta\right)-\frac{g}{\sqrt{2}}e^{\frac{\phi}{2}}\left(i\sigma^2\eta\right)\,, \notag \\
0\,=\,&\frac{g}{\sqrt{2}}\left(e^{-\frac{\phi}{2}}-e^{\frac{\phi}{2}}\right)\eta+\frac{1}{2}g_1^{-1/2}\phi'\left(\sigma^1\eta\right)-\sqrt{2}f^{-1}e^{-\frac{\phi}{2}}b\left(i\sigma^2\eta\right)\,,
\end{align}
where $s\,=\,\pm1$.

\subsection{Supersymmetric solutions}

We present the supersymmetric solutions,
\begin{align}
f\,=&\,\frac{8b^2}{\left(e^{-\frac{\phi}{2}}-e^{\frac{\phi}{2}}\right)^2}\,, \notag \\
g_1\,=&\,\frac{8b^2\left(\phi'\right)^2}{\left(e^{-\frac{\phi}{2}}-e^{\frac{\phi}{2}}\right)^2\left(16g^2b^2-e^{-\phi}\left(e^{-\frac{\phi}{2}}-e^{\frac{\phi}{2}}\right)^2\right)}\,, \notag \\
g_2\,=&\,\frac{8\mathcal{C}^2\left(16g^2b^2-e^{-\phi}\left(e^{-\frac{\phi}{2}}-e^{\frac{\phi}{2}}\right)^2\right)}{\left(e^{-\frac{\phi}{2}}-e^{\frac{\phi}{2}}\right)^2}\,, \notag \\
A_z\,=&\,\mathcal{C}e^{-\phi}\,,
\end{align}
where $\mathcal{C}$ is a constant. Therefore, we have determined all the functions in terms of the scalar field, $\phi(y)$, and its derivative. The solution satisfies all the supersymmetry equations and the equations motion which we present in appendix A. We can determine the scalar field by fixing the ambiguity in reparametrization of $y$ due to the covariance of the supersymmetry equations,
\begin{equation}
\phi(y)\,=\,\log{y}\,,
\end{equation}
where $y\,>\,0$.

Finally, let us summarize the solution. The metric is given by
\begin{equation} \label{metmet3}
ds^2\,=\,\frac{8b^2y}{\left(1-y\right)^2}\left[ds_{AdS_2}^2+\frac{1}{y^2h(y)}dy^2+\frac{\mathcal{C}^2h(y)}{b^2}dz^2\right]\,
\end{equation}
where we define
\begin{equation}
h(y)\,=\,16g^2b^2-y^{-2}\left(1-y\right)^2\,.
\end{equation}
The gauge field is given by
\begin{equation}
\widehat{A}_z\,=\,\frac{\mathcal{C}}{y}\,.
\end{equation}

For the choice of $s=+1$, \eqref{splus}, we find a class of solutions when we have
\begin{equation} \label{regrange3}
y_1<y<y_2\,,
\end{equation}
where $y_1$ and $y_2$ is obtained from $h(y)\,=\,0$,
\begin{equation} \label{rone3}
y_1\,=\,\frac{1}{1-4gb}\,, \qquad y_2\,=\,\frac{1}{1+4gb}\,,
\end{equation}
and $y_1<1<y_2$. We plot a representative solution with $g=1/4$, $b=-0.5$ and $\mathcal{C}\,=\,1$ in Figure 4.

\begin{figure}[t]
\begin{center}
\includegraphics[width=2.0in]{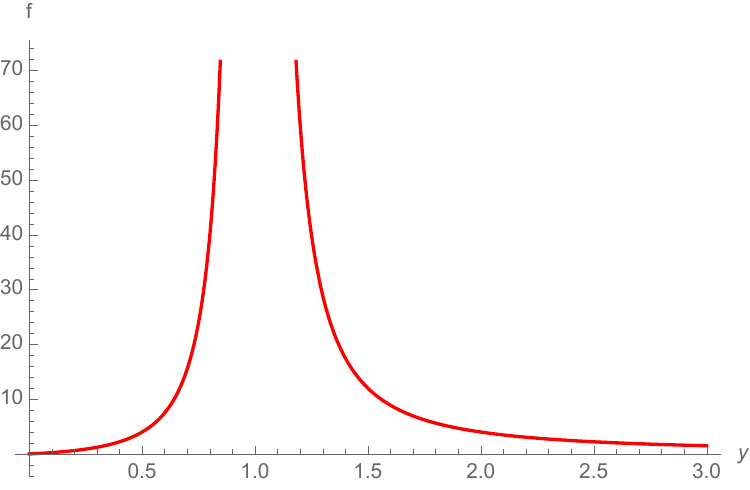} \qquad \includegraphics[width=2.0in]{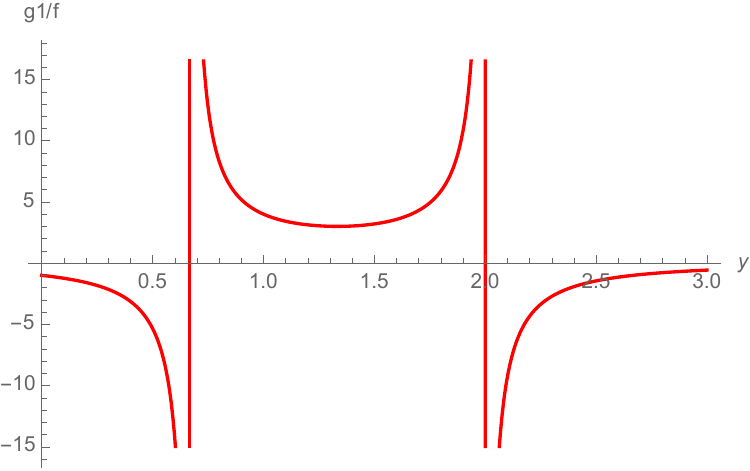} \qquad \includegraphics[width=2.0in]{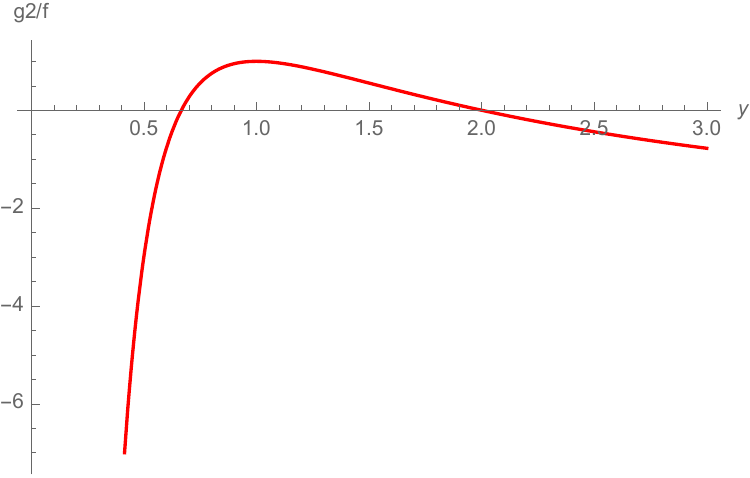}
\caption{{\it A representative solution with $g=1/4$, $b=-0.5$ and $\mathcal{C}\,=\,1$. We consider the solution in the range of $y_1\,=\,0.66\,\,<\,y\,<\,y_2\,=\,2$.}}
\end{center}
\end{figure}

Approaching $y\,=\,y_*$, where $y_*\,=\,y_1\,,y_2$, the metric becomes to be 
\begin{equation}
ds^2\,=\,\frac{8b^2y}{\left(1-y\right)^2}\left[ds_{AdS_2}^2+\frac{4\Big[d\rho^2+\mathcal{C}^2\mathcal{E}_*(b)^2\rho^2dz^2\Big]}{-y_*^2h'(y_*)}\right]\,,
\end{equation}
where we introduced a new parametrization of coordinate, $\rho^2\,=\,|y_*-y|$ and $\mathcal{E}_*(b)$ are
\begin{equation}
\mathcal{E}_1(b)\,=\,\frac{y_1^2h'(y_1)}{2b}\,, \qquad \mathcal{E}_2(b)\,=\,\frac{y_2^2h'(y_2)}{2b}\,,
\end{equation}
for $y=y_1$ and $y=y_2$, respectively. Then, the $\rho$-$z$ surface is locally an $\mathbb{R}^2/\mathbb{Z}_l$ orbifold if we set
\begin{equation}
l_1\,=\,\frac{1}{\mathcal{C}\,\mathcal{E}_1(b)}\,, \qquad l_2\,=\,-\frac{1}{\mathcal{C}\,\mathcal{E}_2(b)}\,,
\end{equation}
where $l_1\,,l_2\,=\,1,\,2,\,3,\ldots\,\,$.

Employing the Gauss-Bonnet theorem, we calculate the Euler characteristic of $\Sigma$, the $y$-$z$ surface, from \eqref{metmet3}. The only contribution to the Euler characteristic is  
\begin{equation}
\chi\left(\Sigma\right)\,=\,\frac{1}{4\pi}\int_\Sigma{R}_\Sigma\text{vol}_\Sigma\,=\,-8g\,\mathcal{C}\,=\,\frac{1}{l_1}+\frac{1}{l_2}\,,
\end{equation}
where $0\,<\,z\,<2\pi$.

The $AdS_2$ solutions we obtain in this section is in the same class of the solutions from Calabi-Yau two-folds in section 4. Although we do not present the calculations, the Bekenstein-Hawking entropy of the solutions diverges at $y\,=\,1$ and is not finite.

\section{Conclusions}

In $U(1)^4$-gauged $\mathcal{N}=2$ supergravity, we constructed supersymmetric $AdS_2$ solutions from M2-branes wrapped on a topological disk in Calabi-Yau two-, three- and four-folds. We uplift the solutions to eleven-dimensional supergravity. For the solutions from topological disk in Calabi-Yau four-folds, the Bekenstein-Hawking entropy is finite and well-defined. On the other hand, from the topological disk in Calabi-Yau two- and three-folds, we could not find solutions with finite Bekenstein-Hawking entropy.

In the construction, we have just considered the three different subtruncations of $U(1)^4$-gauged $\mathcal{N}=2$ supergravity. We would like to construct $AdS_2$ solutions of the full $U(1)^4$ theory. It would generalize the black hole solutions from M2-branes wrapped on a Riemann surface, \cite{Cacciatori:2009iz, Hristov:2010ri, DallAgata:2010ejj}. The Bekenstein-Hawking entropy of these black holes are microscopically counted, \cite{Benini:2015eyy}, by the topologically twisted index, \cite{Benini:2015noa}, of ABJM theory, \cite{Aharony:2008ug}. Therefore, it is an intriguing question to define an index of 3d dual field theory which counts the microstates of our solutions which do not realize the supersymmetry by the topological twist. Furthermore, if it exists, we would like to find the full black hole solution which would interpolate the maximally supersymmetric $AdS_4$ critical point and the $AdS_2$ solution found here. Finally, the canonical form of supersymmetric $AdS_2$ solutions in eleven-dimensional supergravity is presented in \cite{Kim:2006qu} and \cite{MacConamhna:2006nb}. The $AdS_2$ solution we obtain here should fit in the same class.

\bigskip
\leftline{\bf Acknowledgements}
\noindent We would like to thank an anonymous referee for carefully reading and making corrections of the manuscript. This research was supported by the National Research Foundation of Korea under the grant NRF-2019R1I1A1A01060811.

\appendix
\section{The equations of motion}
\renewcommand{\theequation}{A.\arabic{equation}}
\setcounter{equation}{0} 

We present the equations of motion for gauged $\mathcal{N}\,=\,2$ supergravity coupled to three vector multiplets in four dimensions which we review in section 1. The equations of motion are
\begin{align}
R_{\mu\nu}&-\frac{1}{2}\partial_\mu\phi_{12}\partial_\nu\phi_{12}-\frac{1}{2}\partial_\mu\phi_{13}\partial_\nu\phi_{13}-\frac{1}{2}\partial_\mu\phi_{14}\partial_\nu\phi_{14}+2g^2\Big(\cosh\phi_{12}+\cosh\phi_{13}+\cosh\phi_{14}\Big)g_{\mu\nu} \notag \\
&-4e^{\phi_{12}+\phi_{13}+\phi_{14}}\left(F^1_{\mu\rho}F_\nu^{1\rho}-\frac{1}{4}g_{\mu\nu}F^1_{\rho\sigma}F^{1\rho\sigma}\right)-4e^{\phi_{12}-\phi_{13}+\phi_{14}}\left(F^2_{\mu\rho}F_\nu^{2\rho}-\frac{1}{4}g_{\mu\nu}F^1_{\rho\sigma}F^{1\rho\sigma}\right) \notag \\
&-4e^{-\phi_{12}+\phi_{13}-\phi_{14}}\left(F^3_{\mu\rho}F_\nu^{3\rho}-\frac{1}{4}g_{\mu\nu}F^4_{\rho\sigma}F^{4\rho\sigma}\right)-4e^{-\phi_{12}-\phi_{13}+\phi_{14}}\left(F^1_{\mu\rho}F_\nu^{1\rho}-\frac{1}{4}g_{\mu\nu}F^1_{\rho\sigma}F^{1\rho\sigma}\right)\,=\,0,
\end{align}
\begin{align}
\frac{1}{\sqrt{-g}}\partial_\mu\left(\sqrt{-g}g^{\mu\nu}\partial_\nu\phi_{12}\right)+4g^2\sinh\phi_{12}-&2\Big(e^{\phi_{12}+\phi_{13}+\phi_{14}}F^1_{\mu\nu}F^{1\mu\nu}+e^{\phi_{12}-\phi_{13}-\phi_{14}}F^2_{\mu\nu}F^{2\mu\nu} \notag \\
&-e^{-\phi_{12}+\phi_{13}-\phi_{14}}F^3_{\mu\nu}F^{3\mu\nu}-e^{-\phi_{12}-\phi_{13}+\phi_{14}}F^4_{\mu\nu}F^{4\mu\nu}\Big)\,=\,0\,, \notag \\
\frac{1}{\sqrt{-g}}\partial_\mu\left(\sqrt{-g}g^{\mu\nu}\partial_\nu\phi_{13}\right)+4g^2\sinh\phi_{13}-&2\Big(e^{\phi_{12}+\phi_{13}+\phi_{14}}F^1_{\mu\nu}F^{1\mu\nu}-e^{\phi_{12}-\phi_{13}-\phi_{14}}F^2_{\mu\nu}F^{2\mu\nu} \notag \\
&+e^{-\phi_{12}+\phi_{13}-\phi_{14}}F^3_{\mu\nu}F^{3\mu\nu}-e^{-\phi_{12}-\phi_{13}+\phi_{14}}F^4_{\mu\nu}F^{4\mu\nu}\Big)\,=\,0\,, \notag \\
\frac{1}{\sqrt{-g}}\partial_\mu\left(\sqrt{-g}g^{\mu\nu}\partial_\nu\phi_{14}\right)+4g^2\sinh\phi_{14}-&2\Big(e^{\phi_{12}+\phi_{13}+\phi_{14}}F^1_{\mu\nu}F^{1\mu\nu}-e^{\phi_{12}-\phi_{13}-\phi_{14}}F^2_{\mu\nu}F^{2\mu\nu} \notag \\
&-e^{-\phi_{12}+\phi_{13}-\phi_{14}}F^3_{\mu\nu}F^{3\mu\nu}+e^{-\phi_{12}-\phi_{13}+\phi_{14}}F^4_{\mu\nu}F^{4\mu\nu}\Big)\,=\,0\,,
\end{align}
\begin{align}
D_\nu\left(e^{\phi_{12}+\phi_{13}+\phi_{14}}F^{1\nu\mu}\right)\,=\,0\,, \notag \\
D_\nu\left(e^{\phi_{12}-\phi_{13}-\phi_{14}}F^{2\nu\mu}\right)\,=\,0\,, \notag \\
D_\nu\left(e^{-\phi_{12}+\phi_{13}-\phi_{14}}F^{3\nu\mu}\right)\,=\,0\,, \notag \\
D_\nu\left(e^{-\phi_{12}-\phi_{13}+\phi_{14}}F^{4\nu\mu}\right)\,=\,0\,.
\end{align}

\vspace{3.2cm}

\bibliographystyle{JHEP}
\bibliography{20210923}

\end{document}